\documentstyle[12pt,psfig]{article}

\newcommand{\e}{{\rm  e}}

\newcommand{\beq}{ \begin{eqnarray} }
\newcommand{\eeq}{ \end{eqnarray} }
\newcommand{\beqstar}{ \begin{eqnarray*} }
\newcommand{\eeqstar}{ \end{eqnarray*} }

\newcommand{\sla}[1]{\not\!#1}
\newcommand{\E}{ \sla{E} }

\begin{document}
\baselineskip 0.7cm

\begin{titlepage}

\begin{center}

\hfill KEK-TH-625\\
\hfill hep-ph/yymmdd\\
\hfill \today

  {\large  
LEPTON-FLAVOR VIOLATION 
AT FUTURE LEPTON COLLIDERS \\
AND \\
THE ATMOSPHERIC NEUTRINO OSCILLATION\footnote{
Talk given at 34th Rencontres de Moriond: Electroweak Interactions and Unified Theories, Les Arcs, France, 13-20 Mar 1999.}
}
  \vskip 0.5in {\large
    Junji~ Hisano}
\vskip 0.4cm 
{\it 
 Theory Group, KEK, Oho 1-1, Tsukuba, Ibaraki 305-0801, Japan
}
\vskip 0.5in

\abstract {
It can be expected from the result for the atmospheric neutrino by
the SUPERKAMIOKANDE and the CHOOZ result that the the
lepton-flavor violating (LFV) interaction between the second and third
generations exits at the high energy scale.  This leads to a non-vanishing
LFV left-handed slepton mass between the second and third generations,
induced by the radiative correction, in the minimal supergravity
scenario. In this article, assuming that the supersymmetric standard
model with the right-handed neutrinos explains the atmospheric
neutrino result, we show that the reach of the LFV slepton production
in the future lepton colliders can be more significant than that of
$\tau\rightarrow \mu \gamma$.}
\end{center}
\end{titlepage}
\setcounter{footnote}{0}
The lepton-flavor violation (LFV) is one of the characteristic
signatures in the supersymmetric (SUSY) extension of the Standard
Model (SM). Introduction of supersymmetry to the SM is one of the
promising idea beyond the SM, since it is a solution of the
naturalness problem associated with the Higgs boson mass, as
well-known.  In order to make this model phenomenologically viable, we
have to introduce the SUSY breaking terms. This may lead to the LFV by
the gaps between the mass bases of sleptons and of leptons.

In the minimal supergravity scenario, which is the one of the
candidates of the generation mechanism of the SUSY breaking in the
SUSY SM, the magnitude of the LFV depends on the high energy physics
beyond the SUSY SM. In this scenario, the slepton masses are
degenerate and the lepton flavor is conserved at the tree
level. However, if the LFV interaction at the high energy scale, such
as the Yukawa interaction of the right-handed neutrinos, exits, the
radiative correction to the slepton masses is lepton-flavor violating
\cite{BM,HKR}.

The SUPERKAMIOKANDE provided the convincing result for the atmospheric
neutrino anomaly, and showed that it comes from the neutrino
oscillation \cite{superkamiokande}. Combined with the CHOOZ experiment
\cite{chooz}, it is natural to consider that the oscillation is between the
muon and the tau neutrinos, and 
\begin{equation}
\begin{array}{c}
\Delta m^2_{\nu_{\mu} \nu_\tau} \simeq (5\times10^{-4} - 6\times10^{-3})
        {\rm eV^2}, \\
\sin^2 2\theta_{\nu_{\mu} \nu_\tau} >0.82.
\end{array}
\label{atms}
\end{equation}
This means that the LFV interaction, such as the Yukawa
coupling of the right-handed neutrino in the see-saw mechanism
\cite{seesaw}, exits in order to generate the small neutrino masses,
and that the LFV between the second and the third generations in the
slepton mass matrix is generated in the minimal supergravity scenario.

The ways to study the LFV between the second and third generations in
the SUSY SM are {\it i}) the LFV radiative processes such as
$\tau\rightarrow \mu \gamma$ and {\it ii}) the LFV slepton production
processes in the future colliders with the signal $\tau\mu X\E$
\cite{Krasnikov,feng}. In this article we compare the reach of the LFV
slepton production in the future lepton colliders with that of
$\tau\rightarrow \mu \gamma$, assuming that the SUSY SM with the
right-handed neutrinos explains the atmospheric neutrino result. Due
to low statistics of the $\tau\rightarrow \mu \gamma$ experiments and
the weak GIM suppressions in the LFV production processes of slepton,
search for the LFV slepton production in the future lepton colliders
can be more significant than the future experiment of $\tau\rightarrow
\mu \gamma$.

First, we will review the MSSM with the right-handed neutrinos, and
discuss the radiative generation of the LFV in this model. The see-saw
mechanism by introducing the right-handed neutrinos is the simplest
idea to generate the small neutrino masses. The superpotential of the
lepton sector in the SUSY SM with the right-handed neutrinos is
\begin{equation}
W_{\rm MSSM+\nu_R}= 
   f_{\nu_i} U_{Dij}  H_2 N^c_i L_j 
+  f_{l_i} H_1 E^c_i L_j
+  \frac12 M_{ij} N^c_i N^c_j,
\label{eq:superpotential}
\end{equation}
where $L$ is the chiral superfield for the left-handed leptons, and
$N^c$ and $E^c$ are those for the right-handed neutrinos and charged
leptons. $H_1$ and $H_2$ are the Higgs doublets in the MSSM. Here, $i$
and $j$ are generation indices. A unitary matrix $U_D$ is similar to
the Cabibbo-Kobayashi-Maskawa (CKM) matrix in the quark sector. In this 
model the mass matrix for the left-handed neutrinos is given as
\begin{equation}
(m_{\nu_L})_{ij} = U^T_{Dik} (\overline{m}_{\nu_L})_{kl} U_{Dlj}
\label{eq:mnu}
\end{equation}
where $(\overline{m}_{\nu_L})_{ij} = \left[M^{-1}\right]_{ij} f_{\nu_i
} f_{\nu_j}\langle H_2 \rangle^2$. The large mixing angle in the
atmospheric neutrino result (Eq.~\ref{atms}) comes from the unitary matrix
$U_{D32}$ or the neutrino mass matrix $(\overline{m}_{\nu_L})$.  As we
will show later, if $U_{D32}$ is of the order of one, the LFV in the
SUSY SM is enhanced.

The Yukawa interaction of the neutrino given in Eq.~\ref{eq:superpotential}
generates the LFV masses for the left-handed sleptons radiatively \cite{BM}. 
At the logarithmic approximation of one-loop level, the mass matrix
of the left-handed slepton $(m_{\tilde L}^2)$ is given as 
\begin{eqnarray}
(m_{\tilde L}^2) &=&
U_D^\dagger \left(
\begin{array}{ccc}
\overline{m}^2&&\\
&\overline{m}^2&\\
&&\overline{m}^2-\Delta m^2
\end{array}
\right) 
U_D.
\end{eqnarray}
Here, $ \Delta m^2 = \frac{1}{4\pi^2} f_{\nu_\tau}^2 (3 +a_0^2) m_0^2
\log\frac{M_{\rm G}}{M_{\nu_R}}, $ where $m_0^2$ and $a_0$ are the
universal SUSY breaking scalar mass and the SUSY breaking trilinear
coupling in the minimal supergravity scenario. $M_{\rm G}$ and
${M_{\nu_R}}$ are the gravitational and the right-handed neutrino mass
scales. If $f_{\nu_\tau}$ and/or $U_{D32}$ are sufficiently large,
$(m_{\tilde L}^2)_{23}$ is enhanced.\footnote{
The right-handed sleptons cannot get the sizable LFV masses from the
radiative correction in this model since the right-handed leptons are
not coupled with the right-handed neutrinos.
}

\begin{figure}
\begin{center}
\centerline{\psfig{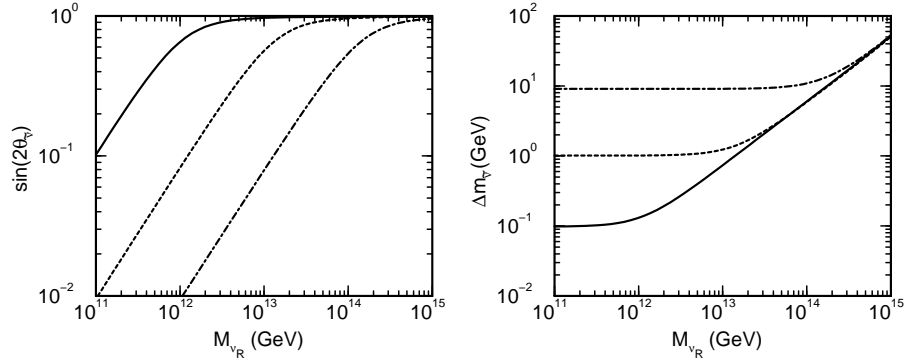}}
\end{center}
\caption{ 
The mass difference $\Delta m_{\tilde{\nu}}$ and the mixing
angle $\sin2\theta_{\tilde{\nu}}$ for the tau and muon sneutrino as a
function of the right-handed neutrino scale in the SUSY SM with the
right-handed neutrino, taking account into the atmospheric neutrino result.
Solid lines, dotted lines, and dash-dotted lines are for $\tan\beta=3$, 10, 30,
respectively.
\label{fig1}}
\end{figure}

The ways to probe $(m_{\tilde L}^2)_{23}$ are
$\tau\rightarrow\mu\gamma$ \cite{hisano} and the LFV production
processes of the left-handed sleptons \cite{hnst}, as we mentioned
before.  The LFV production of the left-handed slepton has statistical
merits. First, this comes from tree-level diagrams while
$\tau\rightarrow\mu\gamma$ is a one-loop process. Second, the GIM
suppression factor in the LFV slepton production is at most $(\Delta
m^2/\overline{m} \Gamma)^2$ with $\Gamma$ the slepton width due to the
slepton oscillation \cite{feng} while the suppression factor in
$\tau\rightarrow\mu\gamma$ is $(\Delta m^2/\overline{m}^2)^2$.  If the
mass difference of the left-handed sleptons is larger than the width,
which is about 1GeV, the suppression factor in the LFV slepton
production is not effective. In Fig.~\ref{fig1} we show the mass
difference $\Delta m_{\tilde{\nu}}$ and the mixing angle
$\sin2\theta_{\tilde{\nu}}$ between the tau and muon sneutrino as a
function of the right-handed neutrino scale $M_{\nu_R}$ in the SUSY SM
with the right-handed neutrinos. Here, we take $m_{\nu_\tau}=7 \times
10^{-2} \ {\rm eV}$ and $U_{D32}=1/\sqrt{2}$.  Also,
${m}_{\tilde{\nu}_{\mu}}=180$ GeV, the wino-like chargino mass 100GeV,
$\tan\beta=3$, 10, 30, and the other parameters are determined in the
minimal supergravity scenario.  If $M_{\nu_R}$ is larger than
10$^{13}$GeV, $\Delta m_{\tilde{\nu}}$ is larger than 1 GeV
\cite{hnst}.

The LFV signals in the direct production of the left-handed sleptons
in the $\e^+\e^-$ collider and the muon collider depend on the mass
spectrum of the SUSY particles. In the minimal supergravity scenario,
the left-handed sleptons and sneutrinos tend to be heavier than the
wino-like chargino or the wino-like neutralino. In this case, the LFV
signal with the largest cross section is
\begin{eqnarray} 
\e^+\e^-(\mu^+\mu^-)&\rightarrow&{\tilde{\nu}}{\tilde{\nu}}^c~{\rm or}~
        {\tilde{l}^+}{\tilde{l}^-}
   \rightarrow\tau\mu +4jets + \E.
\label{signals}
\end{eqnarray}
The jets in the final
states come from the decay of the wino-like chargino and
neutralino. The backgrands from the SM processes are significantly
small since the signal has multi-activities. The most significant
backgrand comes from processes where a tau lepton, which is one of the
tau pair in the tau slepton and sneutrino production, decays
into a muon. In order to reduce this backgrand sufficiently, we need to
enhance the muon and tau lepton identification rates. The energy and impact
parameter cuts for muons are effective for the purpose \cite{hnst}.

\begin{figure}
\begin{center}
\centerline{\psfig{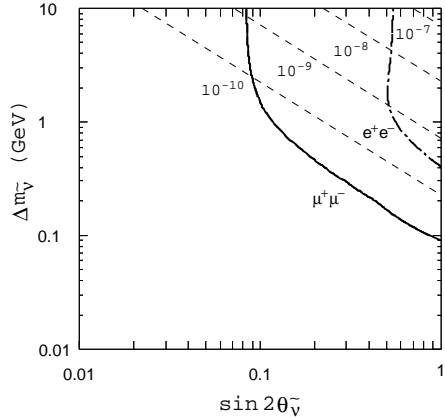}}
\end{center}
\caption{ Significance contours corresponding to $3\sigma$ discovery
as functions of $\sin2\theta_{\tilde{\nu}}$ and $\Delta
m_{{\tilde{\nu}}}$ in the $\e^+\e^-$ collider (solid line) and the
muon collider (dashed-dotted line) with $\sqrt{s} =500$GeV, ${\cal
L}=50fb^{-1}$.  The long dashed lines are for the branching ratios of
$\tau\rightarrow\mu\gamma$ $10^{-7}$, $10^{-8}$, $10^{-9}$, and
$10^{-10}$.
\label{fig2}}
\end{figure}

In Fig.~\ref{fig2} we show the significance contours corresponding to
$3\sigma$ discovery as functions of $\sin2\theta_{\tilde{\nu}}$ and
$\Delta m_{{\tilde{\nu}}}$. The dashed-dot (solid) line is for
$\e^+\e^-$ ($\mu^+\mu^-$) colliders with the center mass energy
500GeV.  We assume the integrated luminosity ${\cal L}=50 {\rm
fb}^{-1}$. For the $\e^+\e^-$ collider we take the impact parameter
cut for muons as $\sigma^{\rm cut}_{IP}= 10 \mu m$.  Here,
$m_{\tilde{\nu}_{\mu}}=180$ GeV, the wino-like chargino mass
100GeV, and $\tan\beta=3$. The other parameters are determined in the
minimal supergravity scenario. The long dashed lines are for the
branching ratio of $\tau\rightarrow\mu\gamma$, $10^{-7}$, $10^{-8}$,
$10^{-9}$, and $10^{-10}$.  Then, the $3\sigma$ significances for
$\e^+\e^-$ collider and the muon collider with the integrated
luminosities ${\cal L}=50fb^{-1}$ correspond to
Br$(\tau\rightarrow\mu\gamma) \sim 10^{-9}$ and $10^{-10}$,
respectively, for $\tan\beta=3$\footnote{
The difference of the significances in the $\e^+\e^-$ collider and the
muon collider comes from a fact that the smuon and muon sneutrino
production cross sections are enhanced by the $t$-channel exchange of
the gauginos in the muon collider. If the integrated luminosity of the
$\e^+\e^-$ collider can reach to 1000$fb^{-1}$, it can compete with
the muon collider with ${\cal L}=50fb^{-1}$.}  \cite{hnst}.

  When $\tan\beta$ is large, Br($\tau\rightarrow\mu\gamma$) is
proportional to $\tan^2\beta$. On the other hand, the LFV slepton
production is not significantly changed, since the mass difference of
slepton is enhanced while the mixing angle is reduced (see Fig.~1). The
future prospect of the reach of Br$(\tau\rightarrow\mu\gamma)$ is at
most $10^{-8}$ at present.  Then, search for the LFV signal in the
slepton production is powerful in the small $\tan\beta$.

In this article, we compared the reach of search for the LFV slepton
production with that of $\tau\rightarrow \mu\gamma$, concentrating on
a case that the LFV mass for the left-handed sleptons $(m_{\tilde
L}^2)_{23}$ is induced in the SUSY SM with the right-handed neutrinos. 
We showed that the search for the LFV slepton production in the future
lepton colliders can be more significant in the LFV study than the
future experiment of $\tau\rightarrow \mu \gamma$.

\newpage
%
%
\def\Journal#1#2#3#4{{#1} {\bf #2}, #3 (#4)}

\def\NCA{\em Nuovo Cimento}
\def\NIM{\em Nucl. Instrum. Methods}
\def\NIMA{{\em Nucl. Instrum. Methods} A}
\def\NPB{{\em Nucl. Phys.} B}
\def\PLB{{\em Phys. Lett.}  B}
\def\PRL{\em Phys. Rev. Lett.}
\def\PRD{{\em Phys. Rev.} D}
\def\ZPC{{\em Z. Phys.} C}
\def\MPL{{\em Mod. Phys. Lett.} A} 

\begin{thebibliography}{99}
\bibitem{BM}
        F.~Borzumati and A.~Masiero,
        \Journal{\PRL}{57}{961}{1986}.

\bibitem{HKR}
        L.~Hall, V.~Kostelecky, and S.~Raby, 
         \Journal{\NPB}{267}{415}{1986}.

\bibitem{superkamiokande}
	Super-Kamiokande Collaboration (Y. Fukuda et al.),
	\Journal{\PRL}{81}{1562}{1998}.

\bibitem{chooz}
	CHOOZ Collaboration,
        \Journal{\PLB}{420}{397}{1998}.

\bibitem{seesaw}
        T. Yanagida,
        in {\sl Proceedings of the Workshop on Unified Theory and
        Baryon Number of the Universe},
        eds. O. Sawada and A. Sugamoto (KEK, 1979) p.95; \\
        M. Gell-Mann, P. Ramond, and R. Slansky,
        in {\sl Supergravity},
        eds. P. van Nieuwenhuizen and D. Freedman
        (North Holland, Amsterdam, 1979).

\bibitem{Krasnikov}
N.V.~Krasnikov,
        \Journal{\MPL}{9}{1994}{791}.

\bibitem{feng}
N.~Arkani-Hamed, H.-C.~Cheng, J.L.~Feng, and L.J.~Hall,
        \Journal{\PRL}{77}{1937}{1996}.

\bibitem{hisano}
        J.~Hisano, T.~Moroi, K.~Tobe, M.~Yamaguchi, and T.~Yanagida,
        \Journal{\PLB}{357}{579}{1995};\\
        J.~Hisano, T.~Moroi, K.~Tobe, and M.~Yamaguchi,
        \Journal{\PRD}{53}{2442}{1996};\\
	J.~Hisano and D.~Nomura,
	\Journal{\PRD}{59}{116003}{1999}.

\bibitem{hnst}
	J.~Hisano, M.M.~Nojiri, Y.~Shimizu, and  M.~ Tanaka,
	hep-ph/9808410.

\end{thebibliography}
\end{document}